\documentclass[runningheads]{llncs}

\usepackage{graphicx}
\usepackage{adjustbox}
\usepackage{tikz}

\usepackage{xspace}
\usepackage{amssymb,amsmath}

\usepackage{times}
\usepackage[utf8]{inputenc}
\usepackage{array,booktabs}
\usepackage{longtable}
\usepackage{latexsym}
\usepackage{enumitem}
\usepackage{bbm}
\usepackage{multirow}
\usepackage{booktabs}

\usepackage{hyperref}
\hypersetup{
    colorlinks=false,
    linkcolor=blue,
    filecolor=magenta,      
    urlcolor=cyan,
    pdftitle={AI Act for the Working Programmer},
    pdfpagemode=FullScreen,
    }

\usepackage[misc]{ifsym}

\newcommand{\takeaway}[1]{\begin{center}
\colorbox{white!90!blue}{\begin{minipage}{0.95\textwidth} \emph{#1}\end{minipage}}
\end{center}}

\def\Art#1{Article~#1~\cite{AIAct}}
\def\Arts#1{Articles~#1~\cite{AIAct}}
\def\Rec#1{Recital~#1~\cite{AIAct}}
\def\Ann#1{Annex~#1~\cite{AIAct}}
\def\fnote#1{\footnote{#1}}

\begin{document}

\title{AI Act for the Working Programmer\thanks{Authors are listed in alphabetic order.
}}

\titlerunning{AI Act for the Working Programmer}
\author{Holger Hermanns \inst{1}
\and  Anne Lauber-Rönsberg\inst{2}  \and \\Philip Meinel\inst{2} \and  Sarah Sterz\inst{1} \and Hanwei  Zhang\inst{1}}
\institute{Saarland University, Saarland Informatics Campus, Saarbrücken, Germany\\
\email{\{hermanns, sterz, zhang\}@depend.uni-saarland.de}
  \and
TU Dresden University of Technology, Institute of International Law, Intellectual Property and Technology Law, Dresden, Germany\\
  \email{\{anne.lauber-roensberg, philip.meinel\}@tu-dresden.de}}

\authorrunning{Hermanns et al.}

\maketitle

\begin{abstract}
   The European AI Act is a new, legally binding instrument that will enforce certain requirements on the development and use of AI technology potentially affecting people in Europe.   
   It can be expected that the stipulations of the Act, in turn,  are going to affect the work of many software engineers, software testers, data engineers, and other professionals across the IT sector in Europe and beyond. The 113 articles, 180 recitals, and 13 annexes that make up the Act cover 144 pages. This paper aims at providing an aid for navigating the Act from the perspective of some professional in the software domain, termed ``the working programmer'', who feels the need to know about the stipulations of the Act.  
\end{abstract}

\setcounter{footnote}{0}

\section{Introduction}
\label{sec:intro}

\raisebox{56.5mm}[0mm][0mm]{{\textcolor{red}{\large \quad\! \textbf{This is a preprint of a paper that will appear in AISoLA 2024.}}}}

After extensive deliberations, the European Union has taken the final
step for adopting the AI Act~\cite{AIAct}. The AI Act aims to ensure the development and deployment of safe and trustworthy AI by relying on a risk-based approach -- the higher the
risks to fundamental rights and society, the stricter the legal
requirements.\footnote{Of course, the AI Act is also not the only law
  that governs the development and use of AI systems. In addition to the
  AI Act, other general or sector-specific laws such as the GDPR, the
  Digital Services Act, anti-discrimination laws, sector-specific
  legislation such as legislation governing medical devices, to name
  just a few, must be observed.} However, the demarcations of the
regulated areas of AI often seem blurred. The idea of this paper is
therefore to provide the ``working programmer''\footnote{In allusion to
  ``ML for the Working Programmer'' by Larry C.~Paulson~\cite{larry}.}
with some initial help in navigating the complexities of the AI Act. In
doing so, we make three main contributions:
\begin{itemize}
\item
  We provide an overview of the regulated AI technologies and how to
  distinguish between them. This is essential for the working programmer
  to determine which legal obligations under the AI Act might apply to
  their work.
\item
  We map the relevant obligations to help the programmer understand
  which parts of the AI Act may be relevant to them. This is supported by a
  flowchart that helps to find the relevant obligations with simple
  questions and to narrow down the complexities of the AI Act.
\item
  Finally, we shed light on the question of in how far programmers can make use of
  ready-made general-purpose AI models, such as large language models, that they want to integrate into their AI
  system. We do this in an interdisciplinary effort of computer science
  and law to help the working programmer understand and anticipate legal
  risks.
\end{itemize}
It appears important to note at this point that there are still many legal uncertainties, as the AI Act is the first of its kind, developed without a blueprint. Most of the legal requirements need further interpretation, which is beyond the scope of this paper. In addition, many requirements in the Act are intentionally broad in order  to be applied to a variety of cases or to be narrowed down by standardisation organisations. In this respect, future work by these organisations is expected to contribute significantly to clarification. Therefore, the goal of this paper can only be to provide the programmer with an overview to enable them to navigate the AI Act effectively.

~

\noindent\textbf{Organization of the paper.}
We start off  by characterising the primary audience for this paper in Section~\ref{sec:addressee}.
We then address the types of AI regulated under the Act in Section~\ref{sec:scope}, providing concrete examples to illustrate various distinctions.
Section~\ref{sec:terri} discusses the scope of the AI Act, including its territorial reach and specific exemptions. 
We then in Section~\ref{sec:gpai-model} move on to the obligations for providers of general-purpose AI models, before we turn to the various obligations for providers of (specific-purpose) AI systems in Section~\ref{sec:ai-systems}.
Section~\ref{sec:ai-on-top-of-gpai},  considers the practical implications of building an AI system on top of a GPAI model.
Finally, we conclude in Section~\ref{sec:conclusion} with a summary of the AI Act's impact on programmers.

\begin{figure}
    \centering  
     
 \adjincludegraphics[width=1.05\textwidth,Clip={.05\width} {.28\height} {.03\width} {.05\height}]{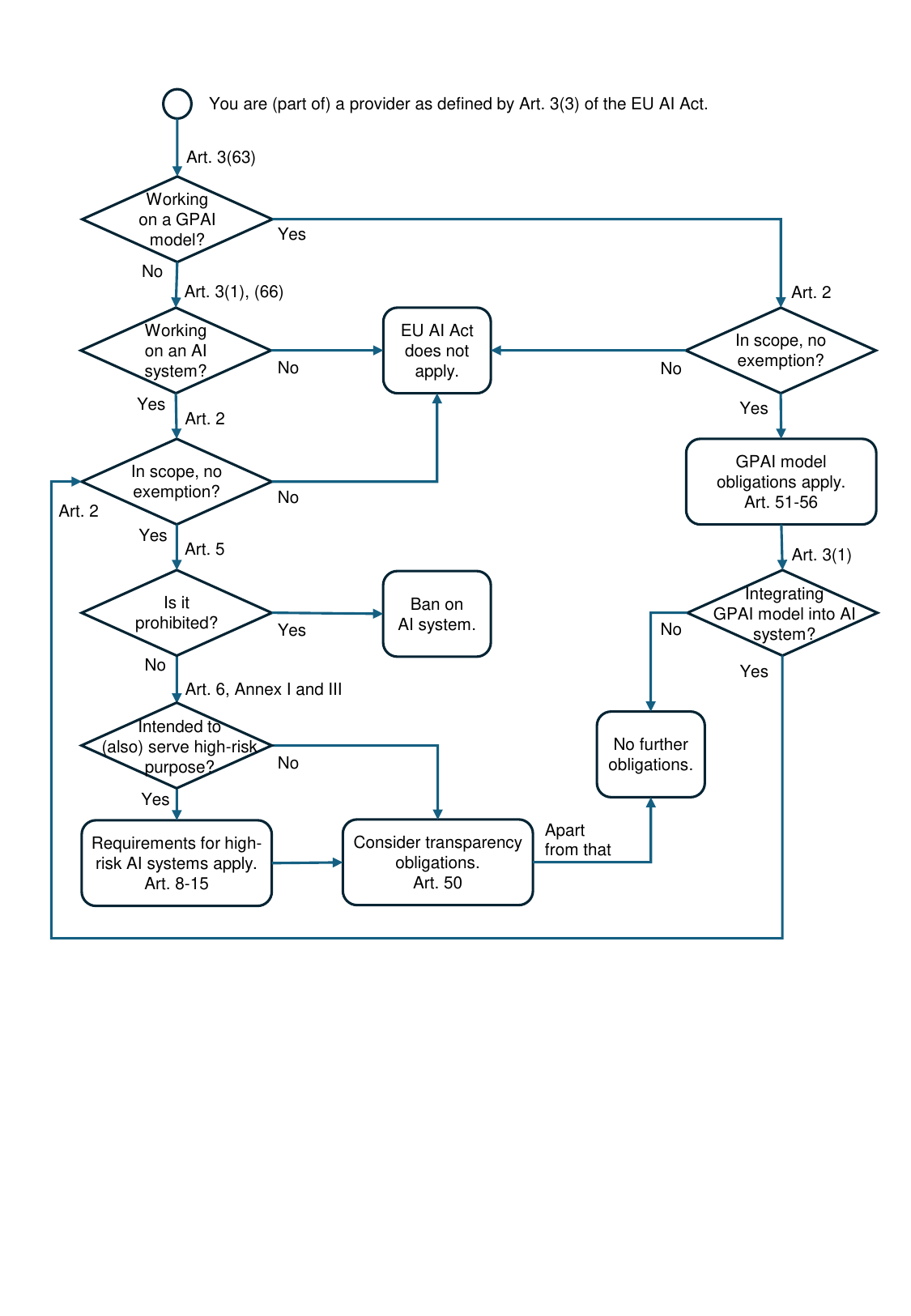}
 
 \raisebox{77mm}[0em][0em]{~\qquad\qquad\qquad\qquad\qquad\qquad\qquad\qquad\qquad\qquad\qquad\qquad\qquad\qquad\qquad\qquad~{\autoref{sec:gpai-model}}\!\!\!\!} 

 \raisebox{11mm}[0em][0em]{~\qquad\qquad\qquad\qquad\qquad\qquad\qquad\qquad\qquad\qquad\qquad\qquad\qquad\qquad\qquad\qquad~{\autoref{sec:ai-on-top-of-gpai}}\!\!\!\!} 
 
\raisebox{65mm}[0em][0em]{{\autoref{sec:ai-systems}}\qquad\qquad\qquad\qquad\qquad\qquad\qquad\qquad\qquad\qquad\qquad\qquad\qquad\qquad\qquad~} 

 \raisebox{124mm}[0em][0em]{{\autoref{sec:scope}}\qquad\qquad\qquad\qquad\qquad\qquad\qquad\qquad\qquad\qquad\qquad\qquad\qquad\qquad\qquad~}

 \raisebox{117mm}[0em][0em]{~\qquad\qquad\qquad\qquad\qquad\qquad\qquad\qquad\qquad\qquad\qquad\qquad\qquad\qquad\qquad\qquad~{\autoref{sec:terri}}\!\!\!\!} 
 
\raisebox{95mm}[0em][0em]{{\autoref{sec:terri}}\qquad\qquad\qquad\qquad\qquad\qquad\qquad\qquad\qquad\qquad\qquad\qquad\qquad\qquad\qquad~} 

 \vspace{-7.5em}
 
    \caption{The working programmer navigating the AI Act}
          \label{fig:enter-label}
\end{figure}

\section{Addressee of the Paper}
\label{sec:addressee}
The particular relevance of the AI Act for the working programmer arises from the fact that it not
only governs the use of AI systems, but primarily sets out requirements
\emph{for their development}. The AI Act addresses a variety of stakeholders
along the AI value chain, among them ``deployer'', ``provider'',
``distributor'', and ``importer''. Against this backdrop, the working
programmer will most likely be considered (being part of) a ``provider''
of AI systems or AI models and this role will therefore be the focus of
the further analysis. The AI Act defines a provider as any natural or
legal person, public authority, agency or other body that develops an AI
system or a general purpose AI model or that has an AI system or a
general purpose AI model developed and places them on the market or puts
the system into service\footnote{Note that `placing on the market' means the
  first making available of an AI system or a general purpose AI model
  on the Union market, \Art{3(9)}. `Putting into service' means
  the supply of an AI system for first use directly to the deployer or
  for own use in the Union for its intended purpose, \Art{3(11)}.} under its own name or trademark, whether for
payment or free of charge.\fnote{\Art{3(3)}.}

In other words, any entity responsible for the development of a system or model 
within the scope of the AI Act could be affected by the requirements of
the Act if it makes the system available to third parties or uses it for
its own purposes.
However, it is important to note at this point, that not only the
development of a new AI system can give rise to the provider's
obligations under the AI Act. Rather, it may be sufficient to
significantly modify an existing system in order to be considered a
provider within the meaning of the AI Act.\footnote{This is the case if
  an AI system is considered to be high-risk after the modifications
  made by the programmer, cf. \Art{25(1) lit. b), c)}. For the
  definition of `substantial modification' see \Art{3(23)}.}

The flowchart in Fig.~\ref{fig:enter-label} takes the programmer's role as a provider under the AI Act as a starting point. The subsequent branches help to determine the resulting obligations by means of yes/no questions. They also reflect the further structure of this paper and can therefore be used as a reading aid.

\takeaway{The usual role of the working programmer facing AI in their daily
work is that of (being part of) a provider in the sense of the AI Act,
and this is what the subsequent discussion presupposes.}

\section{What types of AI are regulated under the AI Act?}
\label{sec:scope}
Compliance with the AI Act is of course only an issue if the AI Act
applies to the projects and tasks of the working programmer.

\subsection{What characterises AI according to the AI Act?}

The definition of ``AI systems'' in \Art{3(1)}, which we will examine in \autoref{sec:ai-systems-def}, contains some insights on what technical approaches the AI Act considers as ``AI'' and which it does not.
As a key characteristic of AI systems, the AI Act underlines their capability to infer how to generate its output, which ``transcends basic data processing by enabling learning, reasoning or modelling''.\fnote{\Rec{12}.} 
The techniques that enable such inference ``include machine learning approaches that learn from data how to achieve certain objectives, and logic- and knowledge-based approaches that infer from encoded knowledge or symbolic representation of the task to be solved.''\fnote{\Rec{12}.} This seems to be what, according to the AI Act, distinguishes AI from ``simpler traditional software systems or programming approaches''.\fnote{\Rec{12}.} The AI Act thus does not apply to ``systems that are based on the rules defined solely by natural persons to automatically execute operations''.\fnote{\Rec{12}.} This means that -- at least to our understanding -- the final version of the AI Act does not 
cover traditional rule-based systems written by humans, even if they are complex and their deployment is associated with high risks.\footnote{A notable exception to this general observation might be expert systems that are not trained, but draw inferences in an elaborate way from an extensive knowledge base, cf. \Rec{12}. However, this will need to be established in future legal interpretation.} This approach has however sparked criticism among legal scholars who advocate a broader scope of application of the AI Act.\footnote{E.g.~Krönke~\cite[p.~529-530]{Kroenke} argues  that it was an open question, to the detriment of legal certainty, whether and from what level more complex rule-based systems also fall under the Regulation's definition of AI. He advocates from a teleological point of view, that the AI Act should also be applicable to such systems, as the use of complex rule-based systems can also lead to risks typical of AI, such as a lack of transparency.} 

   \begin{example}
	The following list contains some concrete and
	distinguishing examples of AI based on our interpretation of the criterion of whether it ``infers, from the input it receives, how to generate outputs''.\footnote{\Art{3(1)}.}
	\begin{enumerate}
		\item \label{compiler} \textbf{Not AI -- a complex but traditional system:} A typical compiler for a high-level
		programming language -- such as Java -- is not considered to be 
		an AI system, regardless of its (potentially excessive) logical complexity. \textit{This system will fall outside the scope of the AI Act, because it does not infer \emph{how} to decide its outputs.}
		 
		 \item\label{refugee-ml} \textbf{AI -- a machine-learned system:} A machine-learned system that decides which refugees should be granted asylum is AI, if the method by which it decides these cases is not directly given in a relevant sense but is machine-learned from data. \textit{This system will fall inside the scope of the AI Act, because it uses inference on \emph{how} to decide its outputs.}

		\item\label{refugee-how} \textbf{AI -- a traditional system that infers how to decide its outputs:} A version of the asylum-system that is not machine-learned but purely logic-based (such as an automated reasoner) is still considered AI, namely if it does not directly infer the output from the input but infers \emph{how to decide} whether an asylum seeker should be granted asylum. So, it first needs to infer a way whereby to arrive at an output for the given input, in this case a way by which to decide for a given asylum case (input) whether the asylum seeker should be granted asylum (output). The output is not given directly, but only after the system applied the method it inferred previously to the input. \textit{This system will fall inside the scope of the AI Act, because it infers \emph{how} to decide its outputs.}
		
		\item\label{refugee-do} \textbf{Not AI -- a traditional system that does \textit{not} infer how decide its outputs:}  If instead the asylum-system is a logic-based system that decides asylum cases by directly inferring the decisions with a human-defined set of rules, it will not be AI. \textit{This system will fall outside the scope of the AI Act, because it does not infer \emph{how} to decide its outputs.}
	\end{enumerate}
The difference between (\ref{refugee-how}) and (\ref{refugee-do}) may seem like a fine line to distinguish at first sight, but it would make the difference between falling under the AI Act and not doing so. These considerations might not be relevant to most practical applications, though, since to us it appears difficult to imagine that purely logic-based systems of the kind discussed in (\ref{refugee-how}) can exist in reality.\footnote{We surveyed a non-representative set of logicians for opinions and the reactions were not supporting the idea that system as in (\ref{refugee-how}) can exist.} Logic-based systems notoriously employ human-made semantics and therefore do not infer how to generate outputs, but instead generate outputs in a direct (though possibly complex) way. Therefore, the system in (\ref{refugee-how}) would be considered to be much like the compiler from (\ref{compiler}).

 Overall, the inference of \emph{how} to arrive at a conclusion seems to be a crucial capability of an AI system according to \Art{3(1)} of the AI Act. Nevertheless, it arguably will be a point of legal and technical controversy in the future. 
This is especially true considering that \Rec{12} names ``logic- and knowledge-based approaches that infer from encoded knowledge or symbolic representation of the task to be solved'' as an example of approaches that enable the kind of inference needed to constitute an AI system.
However, it can broadly be argued that every program ever written uses logic-based approaches that infer from a symbolic representation of the task to be solved. 
So, obviously, a more narrow reading is needed instead. In our opinion, this should pivot on inferences regarding \emph{how} to generate outputs, as explicitly stipulated in \Art{3(1)}. 

This discussion indicates that the recipients of the AI Act are left with uncertainty regarding what is and is not AI according to the Act. However, for many practical cases it will arguably be obvious whether something is AI or not, because most systems in question will use some form of learning from data, and will therefore clearly be AI. This is why many working programmers will probably not have to deal with the above discussed demarcation issues in their work.
\end{example}

\subsection{What different types of AI does the Act regulate?}\label{sec:ai-systems-def}

Within this general classification, the AI Act distinguishes three addressed forms of AI:

\begin{description}
    \item[GPAI models] ``GPAI models'' are defined as AI models (including where such AI models are trained with a large amount of data using self-supervision at scale) that display significant generality, are capable of competently performing a wide range of distinct tasks, and can be integrated into a variety of downstream systems or applications.\fnote{\Art{3(63)}.} The AI Act regards these GPAI models as a fundamental component for subsequent use cases. Therefore, the legal obligations for these GPAI models arise when they are placed on the market, regardless of how this is done, e.g., through libraries, application programming interfaces (APIs), as a direct download, or as a physical copy.\fnote{\Rec{97}.} GPT-4, which serves as the basis for ChatGPT as well as a number of downstream applications, is probably the most relevant example of such a GPAI model.
    \item[AI systems] Secondly, an ``AI system'' is defined as ``a machine-based system that is designed to operate with varying levels of autonomy and that may exhibit adaptiveness after deployment, and that, for explicit or implicit objectives, infers, from the input it receives, how to generate outputs such as predictions, content, recommendations, or decisions that can influence physical or virtual environments''.\fnote{\Art{3(1)}.} With other words, AI systems are the systems that use AI and that may be deployed to end users to achieve explicit or implicit objectives, like the asylum decision system in Example 1. Such AI systems can be either built from scratch or on top of a ready-made GPAI model.
    \item[GPAI systems] In general, AI systems that are built on the basis of ready-made GPAI models, qualify as regular AI systems that serve a set of explicit or implicit objectives. In some cases, however, AI systems that are built on the basis of GPAI models may have the capability to serve a variety of purposes, both for direct use as well as for integration in other AI systems. As a special form of AI systems, the AI Act defines such systems as general-purpose AI systems.\fnote{\Art{3(63)}.} ChatGPT or Google Gemini might be the most prominent examples of such GPAI systems.
\end{description}

\subsection{Relating and explaining the different types of AI}

In order to understand the general scope of the AI Act, it is essential to relate the different forms of AI to each other in light of the drafting process.

\paragraph{The special role of General Purpose AI.}
The AI Act originally only aimed to regulate AI that is capable of performing a distinct range of tasks for a limited number of purposes, e.g. job recruitment decisions or biometric identification systems. However, due to the (then) surprising advent of multimodal large language models in late 2022, spearheaded by the GPT family of models, 
the EU felt pressured to adapt this approach. In an attempt to comprehensively regulate AI along the entire value chain, general-purpose AI was added to the scope of regulation. 
General-purpose AI, as described above, is therefore considered to be a special form of before regulated AI, which were only capable to serve some distinct purposes.

\paragraph{AI systems and AI models.}

Secondly, the AI Act distinguishes between the technical applications of "AI models" and "AI systems," both of which may also serve general purposes. However, the clear distinction between both is controversial, not least because the legislature has failed to explicitly define ``AI model'' in the legal text.\footnote{However, AI models that do not serve general purposes are mentioned in various places in the legal text (cf. \Rec{97}, \Art{3(63)}.} In principle, the AI model can generally be described as algorithms or statistical models that are designed to perform a variety of tasks without being directly usable by end users. It only becomes usable by integrating the AI model into an AI system, which incorporates the AI model and combines it with the components needed to deploy the AI model, such as a user interface.\footnote{\Rec{97}.}
Thus, AI models can be understood as an essential part of the AI system, and AI systems can be understood as final product that may be deployed to end users.
For instance, in summer 2024 the chatbots ChatGPT and Microsoft Copilot would be an AI \emph{systems} wrapping the GPAI \emph{model} GPT-4 of Open AI.

\paragraph{AI Systems based on GPAI Models.}

AI systems built on the basis of ready-made GPAI models will arguably play an important role in the industry. The EU acknowledges this fact, even though AI systems based on GPAI play a less accentuated role in the AI Act. It is difficult to develop a stringent classification of these systems because the AI Act partly relates to the system's \textit{capability} to perform a variety of purposes and partly to the \textit{objectives} pursued by the system as defined by the provider (which might be narrower than the system's capabilities and differ from the purposes pursued by the deployer). 
Accordingly, these AI systems may serve one or more specific objectives, and therefore be considered regular "AI systems" under the AI Act. Possible challenges from this classification will be addressed in Section~\ref{sec:SPAI-GPAI-obligations}. 
In other cases, AI systems that are built on the basis of GPAI models may have the capability to serve a variety of purposes, and therefore qualify as general-purpose AI systems.\fnote{\Art{3(63)}.}
The only obligation imposed not only on providers of AI systems, but also explicitly on providers of GPAI systems, is the transparency obligation under \Art{50(2)} to ensure that synthetic content can be recognized as artificially generated or manipulated. 
Beyond that, it is not entirely clear under which conditions further stipulations by the AI Act apply to such GPAI systems. 
This is particularly relevant for GPAI systems that can be used directly by deployers for at least one purpose that is to be classified as high-risk\fnote{\Art{6}.} or even a prohibited practice \fnote{\Art{5}.}.
These considerations will be the topic of Section~\ref{sec:GPAI-obligations}. 

\paragraph{Introducing the term "SPAI".}
As described above, the AI Act distinguishes AI along two dimensions: AI model or AI system, and its purpose. \autoref{tab:SPAI-GPAI-models-systems} illustrates these categories. To streamline the wording, we will introduce the term \emph{specific-purpose AI} (or \emph{SPAI} for short). We use this to denote all AI that is not general purpose. 
The Act itself does not use this term. However, ``AI system'' is by default meant to encompass both SPAI and GPAI. This is apparent when considering that GPAI systems are defined as a special form of AI systems in \Art{3(66)}.
Moreover, while AI models that do not serve general purposes are mentioned in various places in the legal text (cf. \Rec{97} or \Art{3(63)}), they are not directly subject to obligations under the AI Act. Nevertheless, there may be indirect regulation through the system requirements for high-risk AI systems and the transparency regulation, as they also anticipate development decisions for the underlying models. We therefore consider the term SPAI valuable for understanding the general demarcations of the AI Act.

\setlength\tabcolsep{1em}
\renewcommand{\arraystretch}{1.6}

\begin{table}[t]
\centering

\begin{tabular}{@{}ccc@{}}
\toprule
                             & \textbf{AI models} & \textbf{AI systems} \\ \midrule
\textbf{Specific purpose AI} & SPAI model                                                                & SPAI system                                                               \\
\textbf{General purpose AI}  & GPAI model & GPAI system   \\ \bottomrule                                              
\end{tabular}

\vspace{1em}

\caption{AI models and systems for specific and general purposes.}
\label{tab:SPAI-GPAI-models-systems}
\end{table}

\takeaway{The AI Act addresses AI systems and models capable of inferring how to generate outputs, including machine learning and logic- or knowledge-based approaches, but excludes traditional rule-based systems. It mainly distinguishes between General Purpose AI (GPAI) models, which perform a wide range of tasks, and AI systems, which may or may not use these models for specific or general purposes.}

\section{Scope of Application}
\label{sec:terri}
Other than the regulated forms of AI, the AI Act makes some further stipulations on its scope of application.

\subsection{Territorial Scope}

The territorial scope of application of the AI Act is very broad. Every
AI system and GPAI model that is placed on the market or put into
service within the EU has to comply with the AI Act -- regardless of
whether the provider has its establishment in the EU or is located
there.\fnote{\Art{2(1)(a)}.} This so-called market location principle is well known from other legislations, such as the GDPR. 

Furthermore, the AI Act also applies when not the AI system as such, but only
the output generated by the AI system is used in the EU. The aim of this
provision is to prevent the circumvention of the Regulation and to
ensure an effective protection of persons located in the EU.\fnote{\Art{2(1)(c)}.} For example, when a company established in the EU contracts
services in the field of staff recruitment to an operator established in
a third country, an AI system deployed during this process would need to
comply with the requirements of the AI Act.\fnote{See also \Rec{12}.} For providers in third countries, this means that the AI Act must
also be observed when developing AI systems whose results are to be used
within the EU. However, this presents the challenge that it is necessary
to anticipate during the development of the AI system whether the
outcome will later be used in the EU. In addition, in many cases it may
be difficult to determine whether a particular outcome originates from
an AI system or was generated by humans.

\takeaway{Even if a GPAI model is developed outside the EU, the AI Act
applies to it if it is put on the market or into service there. Even if an AI
system provides its service outside the EU, the AI Act still applies if
that service's output is used in the EU.}

\subsection{Exemptions}

For certain high-risk AI systems in fields such as civil
aviation and motor vehicles, specific provisions have been set out by
other EU legislation. Therefore, the AI Act as such does not apply in these cases, e.g.
to autonomous cars, but instead only the specific rules for the case apply (cf.~\Art{2(2)} and \Ann{I, Section B}). There are also other areas in
which the AI Act does not apply:

\begin{itemize}
\item
  AI systems or AI models, including their output, specifically
  developed and put into service for the sole purpose of scientific
  research and development (\Art{2(6)}).
\item
  Research, testing or development activity regarding AI systems or AI
  models prior to their being placed on the market or put into
  service.\footnote{For some reason GPAI models that are used for
    research, development or prototyping activities before they are
    placed on the market are also explicitly excluded as according to
    \Art{3(63)}.} However, testing in real world conditions shall
  not be covered by that exclusion (\Art{2(8)}).
\item
  AI systems released under free and open-source licences, unless they
  are placed on the market or put into service as high-risk AI systems
  or as an AI system that enables prohibited Artificial Intelligence
  Practices according to \Art{5}. Furthermore, the transparency
  obligations under \Art{50} still apply (\Art{2(12)}).
\item
  Military, defence or national security purposes (\Art{2(3)}).
\item
  Public authorities in a third country and international organisations
  that use AI systems for law enforcement and judicial cooperation,
  provided that such a third country or international organisation
  provides adequate safeguards with respect to the protection of
  fundamental rights and freedoms of individuals (\Art{2(4)}).
\end{itemize}

\takeaway{There are some exemptions in \Art{2} to the applicability of
the AI Act that may be sector-specific or that may relate to the intended uses, such as scientific research and development.}

\section{Requirements for the Provider of GPAI models}
\label{sec:gpai-model}

For developers of GPAI models, the AI Act follows a two-tiered approach,
which is divided into general requirements and additional requirements
for GPAI models with systemic risks. Providers of GPAI models with or
without systemic risks should demonstrate compliance with these
requirements by applying harmonized standards or -- until corresponding
standards have been published -- by complying with codes of
practice.\fnote{\Art{53(4), 55(2), 56}.} The latter are voluntary codes that are meant to be developed with the help from the AI Office, and can be considered as lighter versions of technical standards.

\subsection{General Requirements regarding GPAI models}

\Art{53} primarily contains general obligations regarding the
need for documentation of the GPAI model. The aim of these obligations
is to simplify the use of GPAI models for downstream AI systems. In the
view of the legislature, this requires a good understanding of the models
used in order to enable integration and fulfil the downstream provider's
obligations under the AI Act and other regulations.\fnote{\Rec{101}.} This includes:

\begin{itemize}
\item
  draw up and keep up-to-date the technical documentation of the
  model, including its training and testing process and the results of
  its evaluation;\fnote{\Art{53(1) lit. a)}, this should
    contain, at a minimum, the information set out in \Ann{XI}. This
    technical documentation is intended to be non-public and should only
    be provided, upon request, to the AI Office.}
\item
  draw up, keep up-to-date and make available information and
  documentation to providers of AI systems who intend to integrate the
  general-purpose AI model into their AI systems;\fnote{\Art{53(1)
    lit. b)}, the information provided should enable the
    downstream developers to comply with possible obligations of the AI
    Act, e.g. compliance with high-risk obligations. It should contain,
    at a minimum the information set out in \Ann{XII}.}
\item
  put in place a policy to comply with Union law on copyright and
  related rights and a possible reservation of rights of the copyright
  holders;\fnote{\Art{53(1) lit. c)}.}
\item
  draw up and make publicly available a sufficiently detailed summary
  about the content used for training of the general-purpose AI model,
  according to a template provided by the AI Office.\fnote{\Art{53(1)   lit. d)}.} The aim is to enable authors and other right holders to assess whether their rights have been infringed. The summary should e.g. list the main data collections or sets that went into training the model, such as large private or public databases or data archives, and provide a narrative explanation about other data sources used.\fnote{\Rec{107}.}    
\item
  appoint an authorised representative which is established within the
  Union.\fnote{\Art{54(1)}.}
\end{itemize}

The first two obligations shall not apply to GPAI models that are
released under a free and open-source licence if the model isn't
associated with systemic risks (see next point).\fnote{\Art{53(2)}, this is only the case if the license allows for the access,
  usage, modification, and distribution of the model, and whose
  parameters, including the weights, the information on the model
  architecture, and the information on model usage, are made publicly
  available.} Additionally, information
should be made available to those potential downstream providers who
intend to use the GPAI model for an AI system. In any case, the provider
of the GPAI model should be protected by the recipient's confidentiality
obligations.\fnote{\Art{53(7), 78}.}

\takeaway{\Art{53} imposes documentation and transparency obligations on
general purpose AI model providers in order to facilitate the
integration and compliance of downstream AI systems.}

\subsection{Additional requirements regarding GPAI models with systemic risks}

In case the GPAI model is associated with a ``systemic risk'' the AI Act
imposes some additional obligations on the programmer. According to
\Art{51(1)}, GPAI models are associated with systemic
risk if they either have high-impact capabilities or are considered
equivalent by the Commission. Whether a model has high-impact
capabilities shall be ``evaluated on the basis of appropriate technical
tools and methodologies, including indicators and benchmarks'', and is
presumed when the cumulative amount of computation used for its training
measured in floating point operations is greater than $10^{25}$.\fnote{\Art{51(1) a), (2)}.} If a GPAI model is
associated with systemic risks, the provider is obliged to:

\begin{itemize}
\item
  Perform a model evaluation described in accordance with standardised
  protocols and tools reflecting the state of the art;\fnote{\Art{55(1) lit. a)}.}
\item
  assess and mitigate possible systemic risks at Union level;\fnote{\Art{55(1) lit. b)}.}
\item
  keep track of, document, and report relevant information about serious
  incidents and possible corrective measures to address them;\fnote{\Art{55(1) lit. c)}; ``serious incidents'' are defined in \Art{3(49)}.}
\item
  ensure an adequate level of cybersecurity protection for the
  general-purpose AI model with systemic risk and the physical
  infrastructure of the model.\fnote{\Art{55(1) lit. d)}.}
\end{itemize}

As soon as a GPAI model meets the technical requirements for systemic
risks, the provider must notify the Commission within two weeks at the
latest.\fnote{\Art{52(1)}.} In its notification, the provider
can also put forward arguments that the GPAI model does not present any
systemic risks despite the technical circumstances.\fnote{\Art{52(2)}.}

\takeaway{If a general-purpose AI model poses ``systemic risks'', as defined
by high-impact capabilities, the AI Act additionally requires
standardised evaluations, risk assessments, cybersecurity measures,
incident reporting, and timely notification to the Commission.}

\section{Requirements for the Provider of AI Systems}
\label{sec:ai-systems}

If the developed AI system falls within the scope of application, three
main types of relevant legal ramifications can arise from the AI Act: the system may (1) be prohibited, (2) be considered high-risk, or (3) be none of the two. 

\subsection{Prohibited systems}

The AI Act defines a number of AI systems that are associated with what
it considers to be unacceptable risks. These systems are prohibited
under \Art{5}.\footnote{The systems listed there may be
  already prohibited by data protection law, Union law or the law of the
  Member States, for Union law explicitly \Art{5(8)}.} The catalogue of
AI systems listed here encompasses certain use cases:

\begin{itemize}
\item
  Subliminal techniques that have the objective or effect of materially
  distorting a person or group of persons' behaviour causing them or
  others significant harm.\fnote{\Art{5(1) lit. a)}.} This could
  affect the development of recommendation systems used in social media
  or advertising. The question that will arise here is when a distorting
  effect on one's own behavior reaches a material level. This also
  applies to other systems that exploit any vulnerability of natural
  persons due to their age, disability or a specific social or economic
  situation;\fnote{\Art{5(1) lit. b)}.}
\item
  Systems for the evaluation or classification of natural persons or
  groups based on their social behavior or personality characteristic
  with a social score leading to detrimental or unfavourable
  treatment.\fnote{\Art{5(1) lit. c)}.} This refers to social
  evaluation systems that result in detrimental or unfavorable treatment
  in certain contexts that is either unrelated to the contexts in which
  the data was generated or disproportionate to the behavior of the
  individuals;
\item
  Systems for risk assessment of natural persons to predict the risk of
  a person committing a criminal offence based solely on profiling or
  assessing personality traits and characteristics;\fnote{\Art{5(1)
    lit. d)}.}
\item
  Creating or expanding facial recognition databases through the
  untargeted scraping of facial images from the internet or CCTV
  footage;\fnote{\Art{5(1) lit. e)}.}
\item
  Systems to infer emotions of a natural person in the workplace or
  education institutions except for medical or safety reasons;\fnote{\Art{5(1) lit. f)}.}
\item
  Biometric categorisation systems that categorise individually natural
  persons based on biometric data to infer protected attributes or
  characteristics.\fnote{\Art{5(1) lit. g)}.} This does not apply to
  the labelling or filtering of lawfully acquired biometric datasets or
  categorising of biometric data for law enforcement.
\end{itemize}

The placing on the market, putting into service or use of the
aforementioned AI systems is prohibited. In addition, only the
\emph{use}, but not the development, of ``real-time'' biometric identification systems in publicly
accessible spaces for law enforcement is generally prohibited, yet 
subject to broad exceptions.\fnote{\Art{5(1) lit. h), (2)-(7)}.} 

\takeaway{\Art{5} prohibits AI systems with unacceptable
risks that cause significant harm by distorting behavior, exploiting
vulnerabilities, using social scoring, predicting criminal behavior
based solely on profiling, scraping facial images, inferring emotions
outside of medical or security contexts, or categorizing individuals
based on biometrics, while also restricting real-time biometric
identification in public spaces with exceptions.}

\subsection{High-risk AI systems}
\label{sec:hr}

\begin{figure}
    \centering  
    
 \adjincludegraphics[width=\textwidth,Clip={.05\width} {.74\height} {.03\width} {.0\height}]{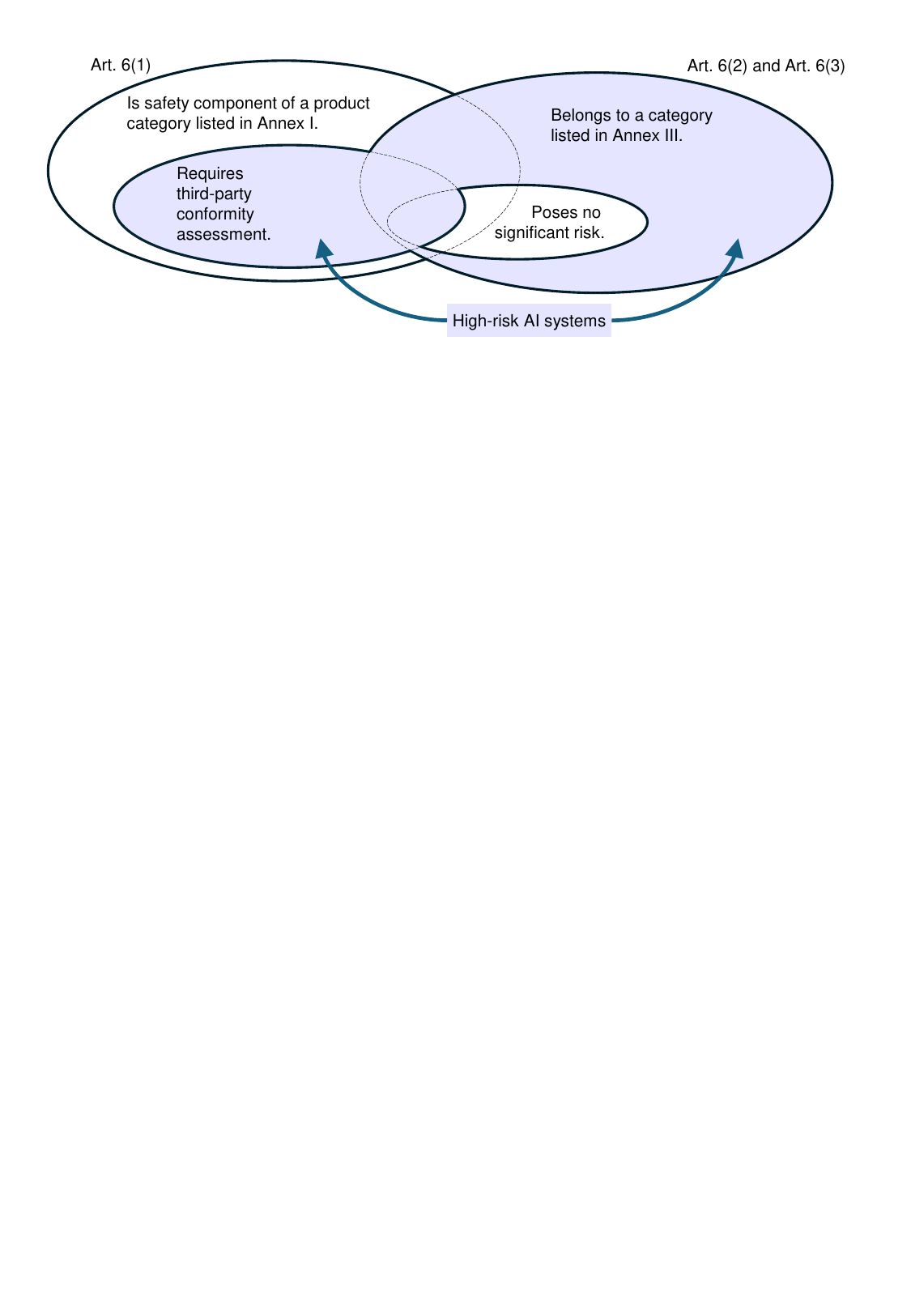}
\caption{Two avenues for an AI system to be classified as high-risk, represented as a Venn diagram.} 
 \label{fig:venn}
\end{figure}

If AI systems are not prohibited by the AI Act, it is generally
permissible to market such systems. However, the AI Act imposes specific
requirements on the development of those AI systems that are deemed to
pose high risks. There are two ways in which an AI system can be
considered high risk (also cf.~\autoref{fig:venn}).

\subsubsection{High-risk AI systems covered by Annex I}

The first option is found in \Art{6(1)}, which states
that the AI system is considered high risk if two conditions are met.
First, the AI system must serve as a safety component of a product
covered by one of the harmonized EU acts listed in \Ann{I}
or be such a product itself.\footnote{``Safety components'' are defined
  in \Art{3(14)} as a component of a product or of an AI system
  which fulfils a safety function for that product or AI system, or the
  failure or malfunctioning of which endangers the health and safety of
  persons or property.} Second, the AI system must be subject to
conformity assessment by a third party with a view to its placing on the
market or putting into service on the basis of those acts listed in
\Ann{I}.

\Ann{I} contains a large collection of EU legislation related to product
safety law, e.g. the Medical Devices Directive, the Machinery Regulation
and the Toys Directive. It is important to emphasise, however, that the
AI system is not automatically considered high risk if the relevant
product falls under one of these legal acts. This is because a
third-party conformity assessment is only required in certain cases
within the respective legal acts. For instance, the above-mentioned
Machinery Directive regulates the safety of machinery placed on the
market. Its scope is therefore quite broad. However, only some, but not all of the
machinery mentioned need to undergo a third party assessment.\fnote{Included are, for example, vehicle servicing lifts or, more relevant, safety components with fully or
partially self-evolving behaviour using machine learning approaches to
ensure safety functions, cf. Art. 25(1), (2) and Annex I Part A (3), (5) of the Machinery Regulation \cite{MachineAct}.} For this reason, the final classification of AI
systems in these cases can only be determined on a case-by-case basis, which will likely require specialized legal expertise.

\takeaway{Under \Art{6(1)}, an AI system is considered
high-risk if it serves as a safety component of a product under
harmonized EU legislation listed in \Ann{I} and requires third-party
conformity assessment for its market placement or putting into service, though
this has to be determined on a case-by-case basis.}

\subsubsection{High Risk Systems covered by Annex III}

Secondly, pursuant to \Art{6(2)} an AI system is considered to be
high-risk if it is listed in \Ann{III}. The classification of these systems is more accessible than the list in \Ann{I} because general areas of application are outlined (points 1-8) and within each of these points specific use cases are listed exhaustively. Accordingly, AI systems used for the following
use cases are generally considered high risk:

\begin{itemize}
\item
  Biometrics, including AI systems used for remote biometric
  identification, biometric categorisation according to sensitive or
  protected attributes or characteristics based on the inference of
  those attributes or characteristics, or emotion recognition;
\item
  AI systems for the use as safety components in the management and
  operation of critical digital infrastructure, road traffic, or in the
  supply of water, gas, heating or electricity;
\item
  Education and vocational training, including systems to determine
  access to institutions, to evaluate learning outcomes, assessing
  appropriate levels of education, or for monitoring students during
  tests;
\item
  Recruitment tools or systems used to make decisions affecting terms of
  employment, like promotion or termination of contractual
  relationships, allocation of tasks, or monitoring;
\item
  Access to and enjoyment of essential private or public services and
  benefits, like healthcare, credits, insurance, or assistance in
  emergencies;
\item
  Law enforcement, e.g. systems to be used to assess the risk of natural
  persons becoming victims of criminal offences, to evaluate the
  reliability of evidence, assess the risk of a person (re-)offending,
  or detect, investigate or prosecute criminal offences;
\item
  Migration, asylum and border control management, e.g. systems to
  assist authorities for the examination of applications, or for border
  control management;
\item
  Administration of justice and democratic processes, including to
  assist a judicial authority in researching and interpreting facts and
  the law or applying the law, or systems to influence the outcome of an
  election or referendum.
\end{itemize}

There is, however, an exception to this general classification of
high-risk AI systems, which was added during the finalisation of the
text of the AI Act: even if the system in question falls under one of
these categories it shall not be considered high-risk if it doesn't
perform profiling of natural persons,\fnote{\Art{6(3) subpara. 3}.} and doesn't pose a significant risk of harm to the health,
safety or fundamental rights of natural persons, including when the system does not
materially influence the outcome of a decision making process.\fnote{\Art{6(3) subpara. 1}.} This applies if the system is intended to
perform a narrow procedural task, to improve the results of prior human
activity, to detect decision-making patterns or deviations therefrom
without replacing human judgement, or to perform a preparatory task to
an assessment related to the use cases in \Ann{III}. This could become an
essential way to avoid the strict obligations that the AI Act defines for
high-risk systems. However, the remaining legal uncertainties associated with the wording of these exceptions will need to be reduced in the future by the deciding courts. 

\takeaway{According to \Art{6(2)}, an AI system is high-risk
if it is listed in \Ann{III} and thus used in sensitive areas such as
biometrics, critical infrastructure, education, recruitment, essential
services, law enforcement, migration or justice, although some
exceptions apply in the absence of significant risk.}

\subsubsection{Obligations: Safety requirements for high-risk AI systems}

Compliance with the following safety requirements contained in \Arts{9 to 15} must be ensured when placing high-risk systems on
the market or putting them into service:

\begin{itemize}
\item
  Establishment, application, documentation and maintenance of a risk
  management system, \Art{9};
\item
  Development and training of high-risk systems with training and
  validation data according to certain quality criteria, data
  governance, \Art{10};
\item
  Technical documentation, \Art{11};
\item
  Automatic recording of processes and events (logging) according to
  recognized standards, \Art{12};
\item
  Transparency and provision of information to the deployer, \Art{13};
\item
  Enabling human oversight during the use of the system, \Art{14};
\item
  Accuracy, robustness and cybersecurity in relation to the intended
  purpose of the specific system, \Art{15}.
\end{itemize}

In general, the system requirements are very broad, which can lead to
difficulties for programmers during implementation. As mentioned above,
technical standards being developed by European standardisation
organisations are foreseen to make the obligations specific and to give clear
guidance to providers. However, it remains to be seen whether these
standards will be sufficient to provide a comprehensive and
workable set of rules.

Compliance with the safety requirements must be demonstrated by the
provider through a conformity assessment procedure.\fnote{\Art{16 lit. f)}.} In most cases this procedure will take place as part
of an internal control. This is particularly the case if a system is
considered a high-risk system within the meaning of \Ann{III} and
available technical standards have been applied completely.\fnote{\Art{43(1), (2)}, \Ann{VI}.} All other high-risk systems must undergo an
external assessment procedure under the above-mentioned harmonized acts,
which will in future also verify compliance with the safety requirements
of the AI Act.\fnote{So-called ``notified bodies'', \Art{43(3)}.}

Furthermore, the provider's responsibility for a high-risk system does
not end with the completion of the assessment procedure. Rather, ongoing
compliance of the system must be ensured. This includes adjustments to
the system if they have reason to consider that the AI system no longer
meets the safety requirements.\fnote{\Art{20(1)}.} In addition,
a post-market monitoring system must be set up and maintained.\fnote{\Art{16 lit. h), 72}.} Finally, the assessment procedure must be
repeated if substantial modifications have been made to the
system.\fnote{\Art{43(4)}; note that self-learning systems do
  not have to go through a new assessment procedure for changes that
  were predetermined by the provider.}

\takeaway{For high-risk AI systems, compliance with \Arts{9 to 15} mandates a comprehensive risk management system, adherence to
quality data standards, technical documentation, automatic logging,
transparency, human oversight, and cybersecurity, with initial and
ongoing conformity assessments required to ensure safety before and
after market entry.}

\subsubsection{Obligations  resulting indirectly from deployers' obligations?} 

Lastly, it appears possible that further obligations of the provider result indirectly from obligations imposed on the deployer.
An example for this phenomenon is rooted in the right to explanation, stipulated in \Art{86}. Deployers of high-risk systems within the meaning of
\Ann{III} must provide clear and meaningful explanations about the role of
the AI system in the decision-making process to affected
persons.\fnote{\Art{86(1)}; ``deployers'' are defined as any
  natural or legal person, public authority, agency or other body using
  an AI system under its authority except where the AI system is used in
  the course of a
  personal non-professional activity, \Art{3(4)}.} Although the
provision is aimed at deployers and its scope is controversial, it may
also indirectly impose obligations on the provider. For example, this
could result in an obligation to at least enable an explanation of the
decisions of an AI system. This does not seem far-fetched, as the AI Act
requires general transparency and explainability of high-risk systems in
various instances.\fnote{\Art{13, 14}.} However, there is currently
very little clarity on what \Art{86} will entail for developers.

\takeaway{Deployers of stand-alone high-risk AI systems must clearly explain
the AI's decision-making role to affected persons, which may indirectly
impose obligations on developers.}

\subsection{Further transparency obligations for certain AI systems}

Lastly, certain AI systems must meet a number of transparency
requirements according to \Art{50}. Transparency in this
context means that programmers must provide specific information to the
end user of their system. This information must be provided to the end
user in a clear and perspicuous manner no later than at the time of the
first interaction or exposure.\fnote{\Art{50(5)}.}

On the one hand, this applies to systems that are designed to interact
directly with end users.\fnote{\Art{50(1)}.} These systems must
be designed and developed to inform end users that they are interacting
with an AI system if this is not obvious to a reasonably well-informed,
observant, and circumspect person.\footnote{Note the exception in \Art{50(1)} for AI systems authorised by law to detect, prevent,
  investigate or prosecute criminal offences.} In other words, chatbots, for example, 
must be designed and developed in such a way that they make it clear to
their users that they are bots (``Bot-Disclosure'').\footnote{Critique to
  this approach, cf.~\cite[p.~106]{VealeZuiderveenBorgesius+2021+97+112}.}

On the other hand, generative AI must make clear the artificial origin
of its output.\fnote{\Art{50(2)}.} This equally applies to AI
systems and GPAI models that generate synthetic audio, image, video or
text content. The providers of these models must ensure that the output
of their systems is labelled in a machine-readable format and can be
identified as artificially generated or manipulated.\footnote{The AI Act
  also names some examples for appropriate techniques in its \Rec{133}, such as watermarks, metadata identifications, cryptographic
  methods for proving provenance and authenticity of content, logging
  methods, fingerprints.} This obligation does not apply if the system
or model makes minor changes to the input data. 
However, it is unclear at what point a change to an input is considered to be minor, so providers would be better advised to introduce a general label for all outputs.

\takeaway{Certain AI systems must meet transparency requirements according
to \Art{50}, necessitating that systems interacting with
users clearly disclose their AI nature and that generative AI outputs
are labelled as artificially created or manipulated.}

\section{Building an AI System on top of a GPAI model} \label{sec:autora}\label{sec:ai-on-top-of-gpai}

We now aim at reflecting some practical consequences of what we discussed thus
far. We consider the question of how to navigate the AI Act if aiming to
build an AI system that uses a GPAI model as one of its core functional
components. These downstream AI systems can take two forms
under the AI Act. They can either be considered SPAI systems that are built on top of a GPAI model if they have specific purposes. Or they can be considered a GPAI system if the final system itself has the capability to serve a variety of purposes.\fnote{Defined in \Art{3(66)}; \Rec{100} makes this clear when stating that ``when a general-purpose AI model is integrated into or forms part of an AI system, this system should be considered to be general-purpose AI system when, due to this integration, this system has the capability to serve a variety of purposes''.}

\subsection{Obligations for SPAI systems built on top of a GPAI model}
\label{sec:SPAI-GPAI-obligations}

When integrating a GPAI model into an SPAI system that is categorised as high-risk, some compliance issues may arise. To illustrate these issues, let us elaborate a use case. Specifically, we consider the
following hypothetical auto-ranking system used by a university
admissions committee to decide which applicant should be granted
admission to a particular degree program:

\begin{example} 
The AI system ``AutoRa'' produces a (partial) order of applicants by
aggregating pairwise decisions of the form ``candidate X is no less
suited than candidate Y''. The resulting partial order is presented to
an admission board, which uses it to decide where to put the bar for
acceptance, by basically cutting the order in two parts: one below the
acceptance threshold (that are rejected), and one above (that are accepted).
The individual decisions ``candidate X is no less suited than candidate
Y'' are relegated to a GPAI model ``SmartLLM'', which AutoRa is built ontop. Among its multitude
of competences, SmartLLM has the ability to directly process the complete
application material of candidate X and of Y (each presented as a single
pdf consisting of CV, motivation letter, reference letters, English
language certificate, transcript of records, and other certificates). It thereby judges who, if any of the candidates, is better suited for the study program.  Prompt engineering has been used to make sure SmartLLM
will answer ``Yes'' or ``No'' according to its judgement to the question ``Is candidate X is less
suited than candidate Y?'' Furthermore, the GPAI model has been fine
tuned by retraining it based on the decisions taken over the past five
years by the board,\footnote{We assume that the applications of earlier
  candidates where graded on a 5-point scale from ``strong reject'' to
  ``strong accept'' by the selection committee, and that the fine tuning
  is done according to these ratings.} using state-of-the-art data
curation techniques. The AI system AutoRa internally uses standard
rules to minimize the number of calls to SmartLLM when assembling the
partial order.  
\end{example}

\paragraph{Legal classification under the AI Act.} AutoRa, being an AI system that decides over the access to higher education, clearly falls under the high-risk systems listed in \Ann{III
3 (a)}. And thus AutoRa needs to fulfil the requirements for high-risk
systems. This includes compliance with the quality requirements for
training data under \Art{10}, for instance with respect to possible biases that are
likely to lead to discrimination,\fnote{\Art{10(2) f)}.} a risk that
is notoriously high for the university admission situation in our focus.
In this regard, \Art{10} requires that the training, validation and testing data sets shall be subject to data governance and management practices appropriate for the intended purpose of the system,\fnote{\Art{10(2)}.} and for this intended purpose it does not make a difference if the underlying technology uses a GPAI model or not. 

\paragraph{Compliance through the training of the underlying GPAI model.} First, one could wonder if the original training of SmartLLM can fulfil these requirements and whether compliance could be reasonably ensured by the provider of AutoRa. We doubt that the training of GPAI models is sufficient for high-risk use cases, as they are usually trained on very large datasets and therefore lack the necessary curation. In any case, however, the providers of downstream AI systems probably won't have the necessary means to verify whether or not the training of the GPAI model suits the intended purposes of their AI system.
As outlined above, the provider of GPAI models, in our case SmartLLM, has certain information
obligations for the benefit of downstream providers who build an AI
system with a specific purpose based on the GPAI model.\fnote{\Art{53(1) lit. b)}.} The provision of this information also is intended to
make it easier for providers of downstream systems to comply with the
obligations of the AI Act,\fnote{\Art{53(1) b) (i)}.} which in our
setting is the provider of AutoRa. However, the
information obligations shall apply without prejudice to the provider's
intellectual property and trade secrets.\fnote{\Art{53(1) b)}.} Accordingly, the
majority of providers of GPAI models will reduce the provided
information to a minimum, thereby hampering verification of compliance through downstream providers. 

We therefore doubt that the information requirements for GPAI model providers are sufficient to enable downstream providers to verify compliance. This is particularly true with regard to the obligations concerning training data, as according to our reading the AI Act stipulates that the provider of AutoRa as a high-risk AI system would have to duly examine the training data in view of possible biases that are likely to lead to prohibited discriminatory output and to take appropriate countermeasures.\fnote{\Art{10(2)(f) and (g)}. See, however, regarding the obligations of the provider of a GPAI model in case of fine-tuning \Rec{109}.}

\paragraph{Compliance through fine-tuning.} Another option to ensure compliance would be to adapt an existing GPAI model sufficiently to
the purpose of the AI system built on top of it through fine-tuning,
like it was done in the case of AutoRa. The AI Act mentions the modification of GPAI models through fine-tuning only in passing: According to \Rec{109}, fine-tuning results in the responsible party entering into the obligations of GPAI model providers with respect to the changes. However, the question of how fine-tuning affects compliance with the requirements for high-risk systems remains unanswered.
This being said, pre-training a general model and
then fine-tuning it on specific downstream tasks has proven to be highly
effective, particularly in natural language processing~\cite{han2024parameter}. The
advancement of LLMs, such as ChatGPT, exemplifies this success~\cite{zhong2023can,yang2023harnessing}. Fine-tuning algorithms can be categorized into additive,
selective, reparameterized, and hybrid fine-tuning based on their
operations~\cite{han2024parameter}. \emph{Additive fine-tuning} maintains the
pre-trained model unchanged and introduces minimal trainable parameters,
such as an adapter layer~\cite{wang2022adamix,pfeiffer2020adapterfusion,houlsby2019parameter} or a soft prompt~\cite{wang2023aprompt,lester2021power,li2021prefix}.
Instead of adding parameters, \emph{Selective fine-tunings} select a
subset of the existing parameters to adapt, enhancing model performance
over downstream tasks~\cite{vucetic2022efficient,ding2023parameter,guo2020parameter}. Reparameterization involves
equivalently transforming a model's architecture from one form to
another by transforming its parameters. Research~\cite{aghajanyan2020intrinsic} shows that
common pre-trained models exhibit exceptionally low intrinsic
dimensionality, making low-dimensional reparameterization effective for
fine-tuning~\cite{aghajanyan2020intrinsic,yang2023bayesian,liu2023moelora,valipour2022dylora}. \emph{Hybrid fine-tunings} combine the
advantages of diverse fine-tuning approaches or analyze the similarities
among them to establish a unified perspective~\cite{hu2023llm,chen2023parameter,mao2021unipelt}.
However, one could doubt that fine-tuning makes the AI system compliant with the requirements for
high-risk systems. For instance, catastrophic forgetting in fine-tuning,
which is directly caused by the inaccessibility of historical data, is a
research problem that has been extensively studied by numerous
researchers~\cite{hayes2020remind,wang2017growing,chen2020recall,cong2020gan}. Determining what information from
the pre-training data is retained or forgotten after fine-tuning is
challenging. For reasons of privacy and fairness, techniques for
deliberately forgetting specific information during fine-tuning have
been proposed~\cite{chundawat2023can,baik2020learning}. Thus, even if a fine-tuned GPAI model meets the
requirements of the AI Act, we cannot directly demonstrate that a
high-risk SPAI system using it will satisfy the data quality requirements for such systems under the AI Act. Additional
techniques may address these issues, but they must be tested
specifically.
Finally, fine-tuning a GPAI model for a specific purpose could also make it more difficult to meet the other system requirements, in particular for accuracy, robustness and cybersecurity under \Art{15}. Though fine-tuning is effective for downstream tasks in
terms of accuracy, its security and robustness are not guaranteed.
Research~\cite{qi2023fine} indicates that simply fine-tuning with benign and
commonly used datasets can inadvertently degrade the safety alignment of
LLMs. Their experimental results demonstrate that fine-tuning LLMs makes
it easier to bypass their safety guardrails, causing the model to
respond to nearly any harmful instructions~\cite{qi2023fine}. Additionally, the
security and robustness of general-purpose AI models, such as LLMs,
remain unresolved challenges requiring further research~\cite{sun2024trustllm,li2024wmdp,zeng2024johnny,chen2023combating}. These issues make it difficult to guarantee the
safety of using GPAI models for high-risk applications.

\paragraph{Upshot of example.} 
The example of AutoRa illustrates the challenges associated with downstream uses of GPAI models in high-risk AI systems. To our understanding, the AI Act obliges the downstream system providers to ensure that the high-risk requirements are met, including with respect to the underlying GPAI model. This may pose a major hurdle for the downstream provider when using a third party's GPAI model. The legal approach chosen by the AI Act thus arguably makes it difficult in practice to integrate GPAI models into SPAI systems
, thereby jeopardising economic efficiency and threatening to stifle innovation. If, on the other hand, European legislation had limited the legal obligations of downstream providers 
to the parts of the AI system they have developed themselves, this would not have met the objective of ensuring a high standard of protection for societal and individual interests. This shows the trade-offs faced by the European legislature when attempting to regulate GPAI models and to create an effective regulation along the AI value chain.

\takeaway{Integrating a GPAI model into a high-risk SPAI system necessitates meeting strict system requirements, particularly concerning the quality of training data. Our case study demonstrated the significant challenges providers face in fulfilling these requirements.}

\subsection{Obligations for GPAI systems}
\label{sec:GPAI-obligations}

GPAI systems (as opposed to GPAI models), on the other hand, are only briefly mentioned in the AI
Act. For example, generative AI based on GPAI systems is subject to a
special transparency obligation under \Art{50(2)}. Beyond that there are no specific requirements for GPAI systems. 
However, it is unclear which other obligations may also apply.

\paragraph{Applying high-risk requirements to GPAI systems.}
In our opinion, the applicable rules depend on the purposes of the GPAI system as designated by the provider.\fnote{\Art{3(12)}.} When a GPAI system is designated to serve a variety of purposes, at least one of which are considered to be, e.g., high-risk, arguably the additional obligations for high-risk AI systems apply. However, when a GPAI system is determined to serve a variety of purposes none of which is categorized as high risk, even though a high-risk use might be factually possible, then additional obligations for high-risk AI systems do not apply. \Art{25(1)(c)} seems to support this reading as it implies that GPAI systems can also be classified as high-risk AI systems.

\paragraph{Repurposing GPAI systems.}
However, this approach can be criticised since it gives the providers the power to determine the applicable rules by defining the purposes of the GPAI system. Because the deployer could then use the GPAI system for different purposes anyway, this could arguably lead to circumventing the requirements for high-risk systems.
The AI Act tries to address this issue in \Art{25(1)(c)} which requires that a deployer who modifies the intended purpose of a GPAI system which has not been classified as high-risk in such a way that the system becomes a high-risk AI system. 
In this case, the deployer turns into the responsible provider and has to make sure that the relevant system requirements are met.\fnote{\Art{25(1)(c)} and \Rec{84}.} We refer to this new provider as a "deployer-provider".
However, even though the AI Act obliges the initial provider of the initially non-high-risk GPAI system to cooperate with the deployer-provider,\fnote{\Art{25(2)} and \Rec{85}. The initial provider ``shall make available the necessary information and provide the reasonably expected technical access and other assistance''.} it is difficult to imagine how the deployer-provider can fulfill all these obligations, e.g. with regard to the quality of training data.\fnote{\Art{10}.} These issues are reflected in our example in Section~\ref{sec:SPAI-GPAI-obligations}, which refers to the related case of using GPAI models inside a high-risk AI system.
Additionally, the obligation to cooperate with the downstream provider does not apply if the original provider of the GPAI system has clearly specified that the system is not to be changed into a high-risk AI system.\fnote{\Art{25(2)}, \Rec{86}.}

\paragraph{Implications for the Provider.}
For the initial provider, this means that they need to consider carefully what purposes they want to set for their GPAI system. In our view, they have three options: Either they commit to complying with the relevant system requirements, thereby enabling end users to use their systems for high-risk purposes. Or, second, they can define the purposes of the GPAI system without matching them with high-risk use cases, and then try to provide due support to downstream users. Third, they could define the purposes without high-risk use cases and specify in the terms of use that the system may not be used for high-risk purposes, thereby avoiding the necessary support. It should be noted, however, that the provider could reduce the attractiveness of its GPAI systems if it excludes high-risk uses in order to circumvent the legal requirements, as this will affect the overall usefulness of its product.

\takeaway{When it comes to classifying a GPAI system as high-risk 
(or not), the AI Act does not focus on the system's capacities, but on its purposes as designated explicitly by the provider
. A deployer using that system for a different purpose than designated 
 becomes a provider as well and has to meet the thus induced legal obligations. Furthermore, obligations for GPAI systems include specific transparency requirements for generative AI.}

\section{Conclusion}
\label{sec:conclusion}
This paper has navigated the AI Act from the perspective of the working programmer and has demonstrated that the AI Act contains numerous obligations of relevance for their daily work. Beyond that, other provisions of the AI Act may have an indirect effect on programmers and their working environment. For example, so-called regulatory sandboxes are to be established by the Member States\fnote{\Art{57}.} that are meant to enable the
testing of AI systems in a controlled environment. Concerning programmers' personal skill set, the AI Act stipulates that providers and deployers have to ensure a sufficient level of ``AI literacy'' of their staff and other persons dealing with the operation and use of AI systems
on their behalf.\fnote{\Art{4}.} AI literacy is defined as the skills, knowledge and understanding that allow providers, deployers and affected persons to make an informed deployment of the AI system, as well as to gain awareness about the opportunities and risks of AI and
possible harm it can cause\fnote{\Art{3(56)}.} and thus goes well beyond a purely technical expertise. So arguably, there will be an obligation for providers and deployers to organise training for staff members within the months to come. 

The AI Act was published in the Official Journal of the EU on 12 July 2024 and thus enters into force on 1 August 2024. However, most parts of the Regulation Act will only become binding two years after. This is meant to give enough time to adapt to the new legal situation. However, some provisions will become applicable earlier, such as the obligation to ensure that staff have sufficient AI literacy (\Art{4}) as well as bans on certain AI practices such as manipulative AI deploying subliminal techniques, social scoring and real-time biometric information systems (\Art{5}). Within 12 months, the provisions requiring Member States to set up authorities as well as procedures on notification of general-purpose AI models presenting systemic risks become applicable (\Art{51}).\fnote{See \Art{113}.} 

Our paper has also shown that there is still a significant degree of legal uncertainties. It is therefore to be hoped that the transition phase of two years will be used by standardisation organisations to develop technical standards as well as by the AI Office to come up with Codes of Conduct \fnote{\Art{95(2) lit.~c)}.} to mitigate insecurities about the legal framework as far as possible. At the same time, this also opens up the opportunity for some working programmers to contribute to the shaping of the future legal framework by submitting comments and engaging in discussions within professional associations, conferences, etc. 

\takeaway{The AI Act has high relevance for the working programmer. This paper constitutes an attempt to pinpont the main facets. However, there is still considerable legal uncertainty.}

\noindent\textbf{Acknowledgements} This work has received financial support by the DFG under grant No.~389792660 as part of TRR~248 -- \href{https://perspicuous-computing.science}{CPEC}, by VolkswagenStiftung as part of Grant AZ 98514 -- \href{https://explainable-intelligent.systems}{EIS},
 and by the Federal Ministry of Education and Research of Germany and Sächsisches Staatsministerium für Wissenschaft Kultur und Tourismus in the program Center of Excellence for AI-research ``Center for Scalable Data Analytics and Artificial Intelligence Dresden/Leipzig'' (ScaDS.AI).

\bibliographystyle{splncs04}
\bibliography{bibliography}

\end{document}